\documentclass{article}
\usepackage{graphicx}
\usepackage{comment}

\usepackage[preprint]{neurips_2025}

\usepackage[utf8]{inputenc} % allow utf-8 input
\usepackage[T1]{fontenc}    % use 8-bit T1 fonts
\usepackage{hyperref}       % hyperlinks
\usepackage{url}            % simple URL typesetting
\usepackage{booktabs}       % professional-quality tables
\usepackage{amsfonts}       % blackboard math symbols
\usepackage{nicefrac}       % compact symbols for 1/2, etc.
\usepackage{microtype}      % microtypography
\usepackage{xcolor}         % colors

\usepackage{pdflscape}
\usepackage{caption}
\usepackage{adjustbox}

\title{CataractSurg-80K: Knowledge-Driven Benchmarking for Structured Reasoning in Ophthalmic Surgery Planning}
\author{%
    Yang Meng$^{1}$ Zewen Pan$^{1}$ Yandi Lu$^{1}$ Ruobing Huang$^{2}$ Yanfeng Liao$^{3}$ Jiarui Yang$^{4}$
}

\begin{document}

\maketitle

\begin{abstract}
Cataract surgery remains one of the most widely performed and effective procedures for vision restoration. Effective surgical planning requires integrating diverse clinical examinations for patient assessment, intraocular lens (IOL) selection, and risk evaluation. Large language models (LLMs) have shown promise in supporting clinical decision-making. However, existing LLMs often lack the domain-specific expertise to interpret heterogeneous ophthalmic data and provide actionable surgical plans. To enhance the model’s ability to interpret heterogeneous ophthalmic reports, we propose a knowledge-driven Multi-Agent System (MAS), where each agent simulates the reasoning process of specialist ophthalmologists, converting raw clinical inputs into structured, actionable summaries in both training and deployment stages. 
%knowledge-driven modality-specific prompts.
Building on MAS, we introduce  CataractSurg-80K, the first large-scale benchmark for cataract surgery planning that incorporates structured clinical reasoning. Each case is annotated with diagnostic questions, expert reasoning chains, and structured surgical recommendations. We further introduce Qwen-CSP, a domain-specialized model built on Qwen-4B, fine-tuned through a multi-stage process tailored for surgical planning. Comprehensive experiments show that Qwen-CSP outperforms strong general-purpose LLMs across multiple metrics. Our work delivers a high-quality dataset, a rigorous benchmark, and a domain-adapted LLM to facilitate future research in medical AI reasoning and decision support.
\end{abstract}

\section{Introduction}

Cataract is the leading cause of blindness globally, and with the accelerating aging of the population, the demand for cataract surgery continues to rise sharply~\citep{davis2016evolution}. Modern cataract surgery critically depends on meticulous preoperative planning, which requires ophthalmologists to integrate heterogeneous examination reports, including optical coherence tomography (OCT), Pentacam imaging, fundus examinations, and other medical reports, into a unified clinical assessment. This process culminates in high-stakes decisions, such as IOL selection and surgical risk assessment, and imposes considerable cognitive and time burdens on clinicians.

Recent advancements in artificial intelligence have shown promising results in ophthalmic diagnosis, particularly in tasks such as disease detection, severity grading, and image interpretation~\citep{lu2018applications,tong2020application}. However, these data-driven AI models require large amounts of labeled data,
which is costly. Trained models may also fail to generalize across data acquired from different imaging devices. LLMs, as a subset of generative AI, have demonstrated significant potential in image interpretation, reasoning, and clinical decision-making~\citep{panagoulias2024evaluating,perera2023large}. While models such as ChatDoctor~\citep{li2023chatdoctor}, PMC-LLaMA~\citep{wu2024pmc}, and Med-PaLM~\citep{tu2024towards} have demonstrated capabilities in medical reasoning and multi-turn dialogue generation, most remain limited to general-purpose QA or static imaging tasks. 

Effectively applying LLMs to ophthalmic surgery demands the ability to integrate highly structured and visual data, adapt to ophthalmology-specific workflows, and generate interpretable, clinically grounded recommendations. Developing models tailored for cataract preoperative planning introduces domain-specific challenges that need to be systematically addressed. 
First, ophthalmic examination reports integrate multimodal data that demand domain-specific expertise for accurate interpretation. General-purpose LLMs often lack the specialized knowledge required to extract clinically meaningful insights from heterogeneous inputs.
Second, large-scale datasets for surgical decision-making in cataract care are still limited. This gap is critical due to the need for transparent, explainable AI to assist in preoperative planning, where nuanced clinical reasoning and risk stratification are crucial. 
In addition, existing models are rarely evaluated in settings that reflect the actual clinical decision-making process, leading to a disconnect between technical performance and practical utility.

% In this paper, we propose a comprehensive framework that addresses the full pipeline of cataract surgical planning using large models. The pipeline of our work is shown in Figure~\ref{fig:Intro}. First, to overcome the limitations of general-purpose LLMs in interpreting multimodal ophthalmic data, we design a knowledge-driven multi-agent system (MAS), where each agent simulates the diagnostic behavior of a specialized ophthalmologist and extracts a clinically meaningful summary aligned with real-world diagnostic standards.
% Building on the MAS framework, we process raw clinical data from cataract patients to construct the first structured dataset-CataractSurg-80K for cataract preoperative planning. Each case includes diagnostic-style questions, step-by-step Chain-of-Thought (CoT) annotations, and structured IOL recommendations. Additionally, we establish the first benchmark for cataract preoperative planning and introduce Qwen-CSP, a domain-specialized model obtained through a multi-stage fine-tuning strategy on Qwen-4B. This fine-tuning process enables the model to deliver robust and interpretable decision support in real-world clinical settings. The comprehensive experiment in multiple tasks, including information extraction, surgical recommendation accuracy, and reasoning quality, has shown that our Qwen-CSP outperforms general-purpose LLMs across multiple metrics. In summary, the main contributions of this paper include:
In this paper, we introduce a comprehensive and systematic framework for AI-driven cataract surgical planning that delivers three core innovations: a knowledge-driven multi-agent reasoning architecture, the first large-scale structured benchmark for ophthalmic surgical reasoning, and a domain-specialized large language model tailored for clinical decision support. Our approach not only bridges the gap between general-purpose LLMs and real-world clinical workflows but also establishes a reproducible pipeline for transparent, structured, and high-fidelity surgical recommendations. The main contributions of this work are as follows:

\begin{itemize}
    % \item To our knowledge, our proposed CataractSurg-80K is the first benchmark aimed at systematically evaluating the capabilities of LLMs in supporting cataract preoperative planning. %Each case includes diagnostic-style questions, step-by-step Chain-of-Thought (CoT) annotations, and structured IOL recommendations,
    % Reasoning-enhanced dataset explicitly modeling clinical reasoning and decision-making logic.
    \item \textbf{Knowledge-driven multi-agent reasoning architecture.} We propose a novel MAS that, for the first time, decomposes ophthalmic surgical planning into collaborative specialist agents, each simulating the diagnostic workflow of an expert ophthalmologist. This modular, prompt-driven framework enables interpretable, clinically aligned extraction and consolidation of heterogeneous examination data, facilitating structured and traceable reasoning processes that go beyond prior single-agent or black-box LLM approaches.
    
    % \item We develop a knowledge-driven MAS that simulates the reasoning process of ophthalmologists to interpret heterogeneous ophthalmic reports and extract clinically relevant information essential for surgical planning.
    \item \textbf{CataractSurg-80K: the first benchmark for structured ophthalmic surgical reasoning.} We construct CataractSurg-80K, the first large-scale, expert-annotated dataset specifically designed for benchmarking structured clinical reasoning in cataract surgery. Each sample systematically models the full decision-making process—consisting of diagnostic-style questions, multi-step chain-of-thought (CoT) reasoning, and structured IOL recommendations—thereby supporting fine-grained supervision and evaluation for both reasoning quality and clinical decision support.

    % \item We propose Qwen-CSP, a domain-specialized model trained using a multi-stage fine-tuning strategy. Comprehensive experiments demonstrate its potential in assisting clinicians with cataract surgical planning by generating structured, clinically aligned recommendations.
    \item \textbf{Qwen-CSP: a domain-specialized LLM for cataract surgical planning.} We introduce Qwen-CSP, a large language model optimized for ophthalmic surgical reasoning via a multi-stage fine-tuning paradigm that successively injects general, medical, and domain-specific knowledge. Qwen-CSP achieves state-of-the-art performance across information extraction, reasoning, and recommendation tasks on CataractSurg-80K, substantially outperforming general-purpose LLMs. Our open-source training pipeline provides a reproducible and extensible foundation for future medical AI research.
\end{itemize}
%Our dataset, benchmark, and model offer a systematic foundation for future research in medical AI reasoning and decision support for cataract surgical planning.

%1. \textbf{We design a knowledge-driven Multi-Agent System (MAS)} capable of parsing and interpreting heterogeneous ophthalmic reports. 

%2. \textbf{We construct \textit{CataractSurg-80K}}, the first large-scale reasoning-enhanced dataset for cataract surgery. Each case includes diagnostic-style questions, step-by-step Chain-of-Thought (CoT) annotations, and structured IOL recommendations, explicitly modeling clinical reasoning and decision-making logic.

%3. \textbf{We develop a multi-stage fine-tuning strategy} that combines general-domain reasoning, biomedical adaptation, and ophthalmic specialization, enabling large models to perform robust, interpretable decision support in real clinical scenarios.

%We evaluate our system across multiple dimensions—including information extraction, surgical recommendation accuracy, and reasoning quality—demonstrating consistent improvements over general-purpose and biomedical LLMs. 

\section{Related Work}

\subsection{Large Language Models in Reasoning}

LLMs have rapidly progressed in few-shot learning, dialogue generation, and reasoning. Models such as GPT-3.5/4 \citep{floridi2020gpt, openai2023gpt}, PaLM \citep{chowdhery2023palm}, and LLaMA \citep{touvron2023llama} demonstrate strong CoT capabilities \citep{wei2022chain, press2022measuring}, laying the foundation for domain-specific adaptations. Biomedical variants like BioGPT \citep{luo2022biogpt} and Med-PaLM~\citep{tu2024towards} extend these models to extract and synthesize clinical knowledge. However, general-purpose LLMs still struggle with tasks requiring structured reasoning over multimodal inputs—such as those seen in surgical planning. To address this, methods like ReAct \citep{yao2023react} and AgentBench ~\citep{liu2023agentbench} simulate specialist workflows via agent-based architectures, though limitations remain in domain-specific multimodal reasoning and structured decision outputs in clinical contexts.

\subsection{Large Language Models in Medicine}

Medical LLMs such as ChatDoctor \citep{li2023chatdoctor}, Med-Alpaca \citep{medalpaca}, and PMC-LLaMA \citep{wu2024pmc} have shown strong performance in biomedical question answering and dialogue generation through instruction-tuning. However, these models cannot generally integrate structured EHR data and diagnostic images, essential for clinical decision-making. Recent multimodal models like Flamingo \citep{alayrac2022flamingo}, CLIP \citep{radford2021clip}, and LLaVA-Med \citep{li2023llava} have extended LLMs to handle medical visual inputs, with promising results in radiology and pathology. Yet, surveys \citep{biomedicalvqa, joshi2023multimodal,ben2019vqa} note that current systems still struggle with domain-specific reasoning, structured output generation, and specialty areas like ophthalmology that demand precise clinical logic.

\subsection{Benchmarks for Evaluating Large Language Models in Medicine}

Recent efforts in benchmarking medical LLMs and vision-language systems have laid important groundwork for evaluating clinical AI systems. Current benchmarks for medical LLMs, such as MedQA~\citep{jin2020disease}, CMB~\citep{wang2023cmb}, and CliMedBench~\citep{ouyang2024climedbench}, primarily evaluate factual recall using multiple-choice QA formats. Medical Visual Question Answering (VQA) iassess the ability to interpret medical images by responding to natural language queries about X-rays, MRIs, and CT scans.~\citep{zhang2023pmc,lau2018dataset,ben2019vqa}
. These approaches have advanced the understanding of single-modality reasoning and language-image alignment. However, current datasets often focus on describing text and image content rather than generating structured outputs needed for clinical decision support, such as treatment recommendations or risk stratification. These limitations call for the development of more comprehensive benchmarks capable of evaluating collaborative, decision-oriented capabilities in medical AI systems.

\section{Methodology}
\label{methodology}
\begin{figure*}[htbp]
\vspace{-1ex}
    \centering
    \includegraphics[width=\linewidth]{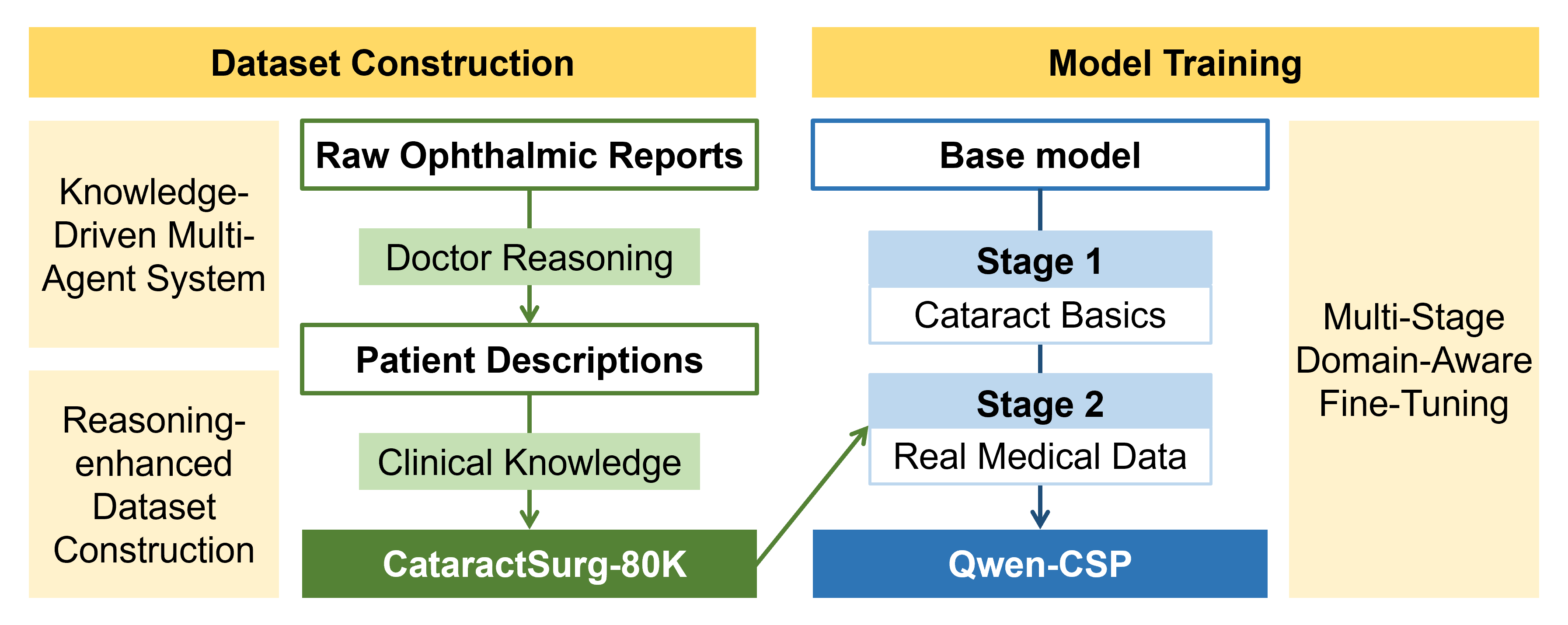}
    \vspace{-2ex}
    \caption{Overview of the multi-agent framework for comprehensive multi-source medical report understanding. The left section illustrates the database construction process, where raw patient reports are processed by collaborative agents to create an inference-augmented dataset. The right section depicts the progressive fine-tuning of the foundational language model using cataract-specific knowledge and real medical data to optimize clinical performance.}
    \vspace{-1ex}
    \label{fig:Intro}
\end{figure*}
% To develop a preoperative decision-making assistant for cataract surgery that enhances precision and efficiency, we propose a comprehensive framework comprising three core modules:  multi-agent extraction module for heterogeneous clinical data processing, a reasoning-enhanced dataset (CataractSurg-80K), and a two-phase training strategy combining pretraining and domain-specific fine-tuning. This framework is designed to systematically analyze multimodal patient data, generate evidence-based surgical recommendations with clinical interpretability, and demonstrate superior robustness in the selection of intraocular lenses (IOL) and the identification of contraindications.

\subsection{Knowledge-Driven Multi-Agent System for Heterogeneous Ophthalmic Reports Interpretation}
% To effectively interpret diverse ophthalmic reports, we propose a collaborative Multi-Agent System (MAS) in which each agent simulates the expertise of a specialized ophthalmologist. Current large language models (LLMs) often lack the specialized knowledge required for interpreting ophthalmic images and associated text. To align the report interpretation process of the LLM with the reasoning methods of real-world ophthalmologist, we develop tailored prompts for each agent, crafted by experienced ophthalmologists. These knowledge-driven prompts can ensure the accurate extraction of clinically significant parameters that carry high medical relevance. Detailed strategies for prompting are available in Appendix A.
To address the limitations of conventional LLMs in clinical data extraction, such as hallucination, lack of transparency, and poor adaptation to heterogeneous medical inputs, we propose a knowledge-driven Multi-agents system (MAS). In our paradigm, each agent is meticulously designed to emulate a specific ophthalmic subspecialty's diagnostic reasoning and data extraction workflow (e.g., OCT, corneal topography, fundus imaging, or master biometric summaries). Unlike generic LLM pipelines, which rely on monolithic, end-to-end extraction and are prone to over-interpretation or privacy risks, our MAS leverages hierarchical, domain-specialized prompting rooted in clinical best practices. Each agent operates with tailored, expert-crafted instructions to extract only those parameters that are clinically verifiable and relevant, strictly avoiding the generation of unsupported or fabricated content.

Each specialist agent executes an independent, modality-aware extraction protocol.The OCT agent uses the same retinal pathology criteria as real ophthalmologists, identifying only observable abnormalities while avoiding any speculative diagnoses. The master reporting agent reconstructs structured biometric data using explicit privacy controls. All agents are constrained by a shared output schema, ensuring that their outputs are both interoperable and traceable, and that each extracted fact is directly associated with its source and the responsible agent. This design not only ensures privacy protection and authenticity, but also enables auditability and clinical trust by providing a transparent chain of reasoning from raw input to structured summary.

% Prompts are structured in a hierarchical manner to align with the specific characteristics of each data modality and its diagnostic focus. For instance, the agent responsible for the OCT report does not simply label presence or absence of fluid but applies ophthalmologist-driven visual criteria to isolate macular abnormalities—such as retinal layer disruption or pigment epithelial detachment—while strictly avoiding over-interpretation beyond visible evidence. This ensures that only clinically verifiable information is captured, consistent with professional reporting standards. Similarly, the agent analyzing the Pentacam corneal topography employs geometric interpretation logic rooted in expert knowledge: it evaluates corneal astigmatism regularity through the angular relationship between principal meridians, and categorizes orientation based on steep axis position. The Master Report Agent reconstructs a highly structured biometric summary, emphasizing bilateral IOL selection logic and emmetropic power extraction while respecting privacy constraints. The Fundus Agent performs a hierarchical classification to determine image type, distinguishing between anterior segment images, standard fundus images, and ultra-widefield modalities.

\begin{minipage}{0.6\textwidth}
  \centering
  \includegraphics[width=1.05\textwidth]{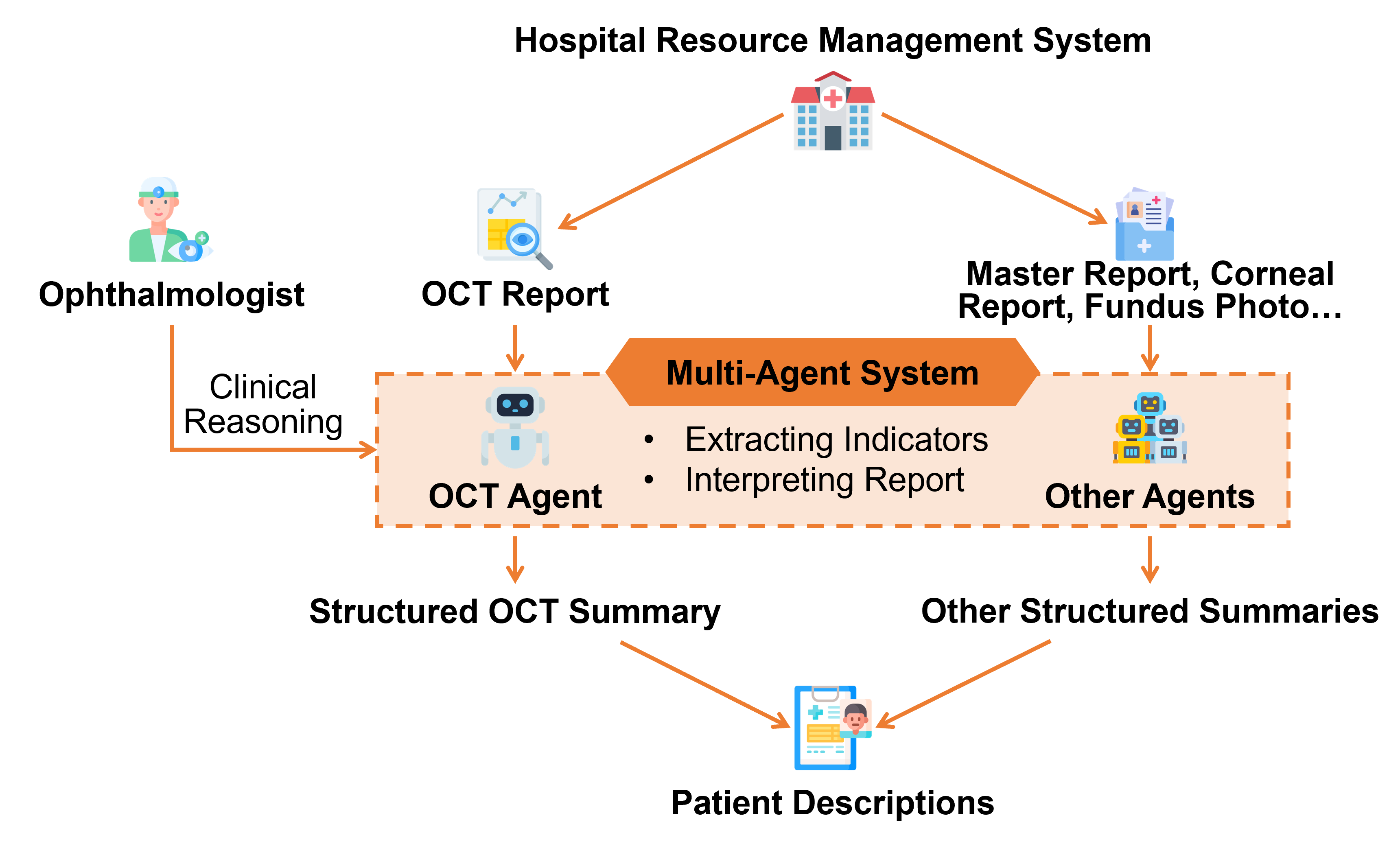} 
  \captionof{figure}{Overview of MAS Architecture} 
  \label{fig:3.1}
\end{minipage}
\hfill
\begin{minipage}{0.34\textwidth}
 % Each agent functions autonomously but adheres to a shared output specification, allowing the MAS to consolidate findings into a unified, structured textual summary. This modular design supports scalability to additional imaging modalities and promotes interpretability by maintaining strict traceability from data source to final output. An overview of the MAS architecture and the agent coordination flow is illustrated in Figure \ref{fig:3.1}.
 Raw ophthalmic reports from the hospital resource management system are first routed to specialized agents, each responsible for interpreting a specific report type using domain-informed criteria. All agents conform to a unified output schema, enabling seamless aggregation and synthesis of their findings into comprehensive patient descriptions. An overview of the MAS architecture and the agent coordination flow is illustrated in Figure \ref{fig:3.1}.
\end{minipage}

\subsection{Reasoning-enhanced Dataset Construction}
\begin{minipage}{0.42\textwidth}
Figure~\ref{fig:3.2} illustrates the construction of the CataractSurg-80K dataset, which bridges raw patient records and actionable clinical recommendations through a multi-stage reasoning framework. Starting from structured patient descriptions, the process unfolds through successive layers of clinical questioning, expert reasoning, and decision synthesis. Each component is tightly coupled with real clinical workflows.
\end{minipage}
\hfill
\begin{minipage}{0.5\textwidth}
  \centering
  \includegraphics[width=1.02\textwidth]{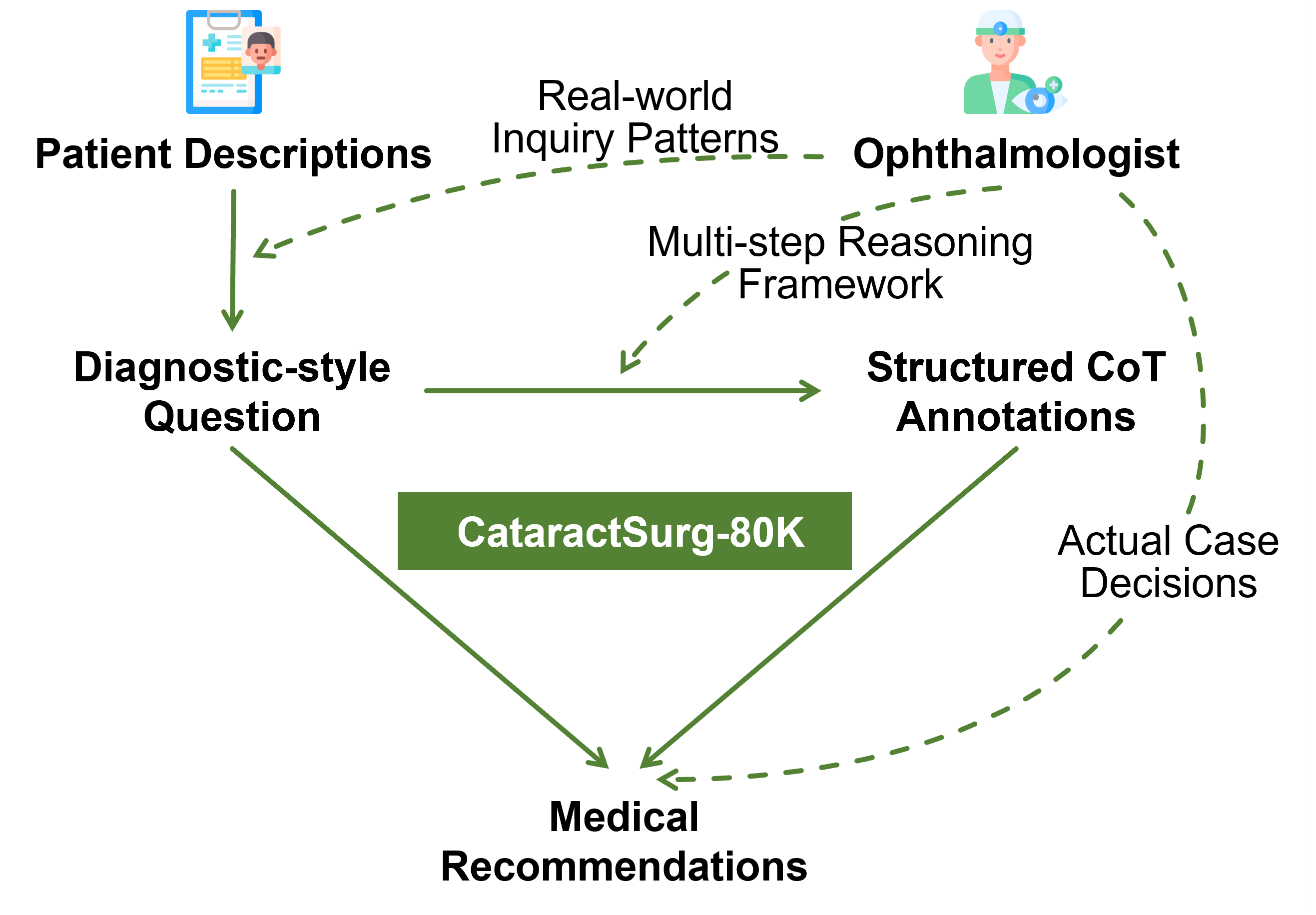} 
  \captionof{figure}{The Construction of CataractSurg-80K} 
  \label{fig:3.2}
\end{minipage}

\textbf{Question Generation.}
To simulate the initial step of clinical decision-making and provide structured input for downstream reasoning and response generation, this stage transforms multimodal patient summaries into concise diagnostic-style questions. These questions are designed to highlight clinically relevant findings, such as axial length, lens opacity grade, corneal astigmatism, and biometric anomalies, without introducing premature diagnosis or treatment bias. The transformation process is guided by prompts co-designed with ophthalmologists, ensuring alignment with real-world inquiry patterns used in clinical consultations. As a result, the generated questions serve as an interpretable and domain-relevant anchor for both reasoning training and surgical plan generation.

\textbf{Chain-of-Thought Annotation.}
To enhance the model's interpretability and emulate expert reasoning pathways, this stage introduces structured CoT annotations for each diagnostic question. Each annotation is generated through an eight-step reasoning framework that covers biometric parameter analysis, corneal topography interpretation, identification of abnormal findings, IOL selection logic, surgical risk assessment, and patient-specific adaptations. The design of this framework is informed by ophthalmologists’ clinical workflows and emphasizes the traceable derivation of each decision from observed data. By aligning annotations with ophthalmologists' decision-making logic, the resulting \texttt{<Question, Complex\_CoT>} pairs serve as an explicit supervisory signal for training interpretable reasoning processes in surgical planning.

\textbf{Medical Recommendation Synthesis.}
To produce actionable, patient-communicable medical recommendations grounded in clinical reasoning, this stage generates structured responses that synthesize the diagnostic question, reasoning path, and physician-aligned recommendation. Each recommendation includes a summary of key findings, personalized IOL selection with type and power, surgical technique suggestions, risk alerts, and postoperative management plans. The generation of recommendation is guided by prompts that require the output to align with actual physician decisions while remaining understandable to patients. These  \texttt{<Question, Response>} pairs support training on full-scenario recommendation generation, ensuring the model can deliver coherent and clinically appropriate plans in real-world deployment contexts.

The overall construction process of CataractSurg-80K explicitly separates reasoning supervision and medical recommendation generation. By combining \texttt{<Question, Complex\_CoT>} and \texttt{<Question, Response>} pairs in a structured manner, our dataset supports fine-grained control over model interpretability and recommendation quality.

\subsection{Multi-Stage Domain-Aware Fine-Tuning for Ophthalmic Surgical Reasoning}

We applied the general language model Qwen3-4B to cataract surgery planning in three steps. First, we fine-tuned it on a cataract-related knowledge dataset to enable it to better understand ophthalmic terms and concepts. Then, we trained it on \texttt{<Question, Complex\_CoT>} pairs from CataractSurg-80K to help the model learn how ophthalmologists reason through real clinical cases. Finally, we further fine-tuned the model on  \texttt{<Question, Response>} pairs to enable it to provide practical surgical recommendations based on real data. This step-by-step process enables the model to accumulate medical knowledge, reasoning ability, and the ability to make clinical recommendations.

\section{Experiments} \label{experiment}
\subsection{Benchmark Dataset Description}
\begin{figure*}[htbp]
\vspace{-1ex}
    \centering
    \includegraphics[width=1\linewidth]{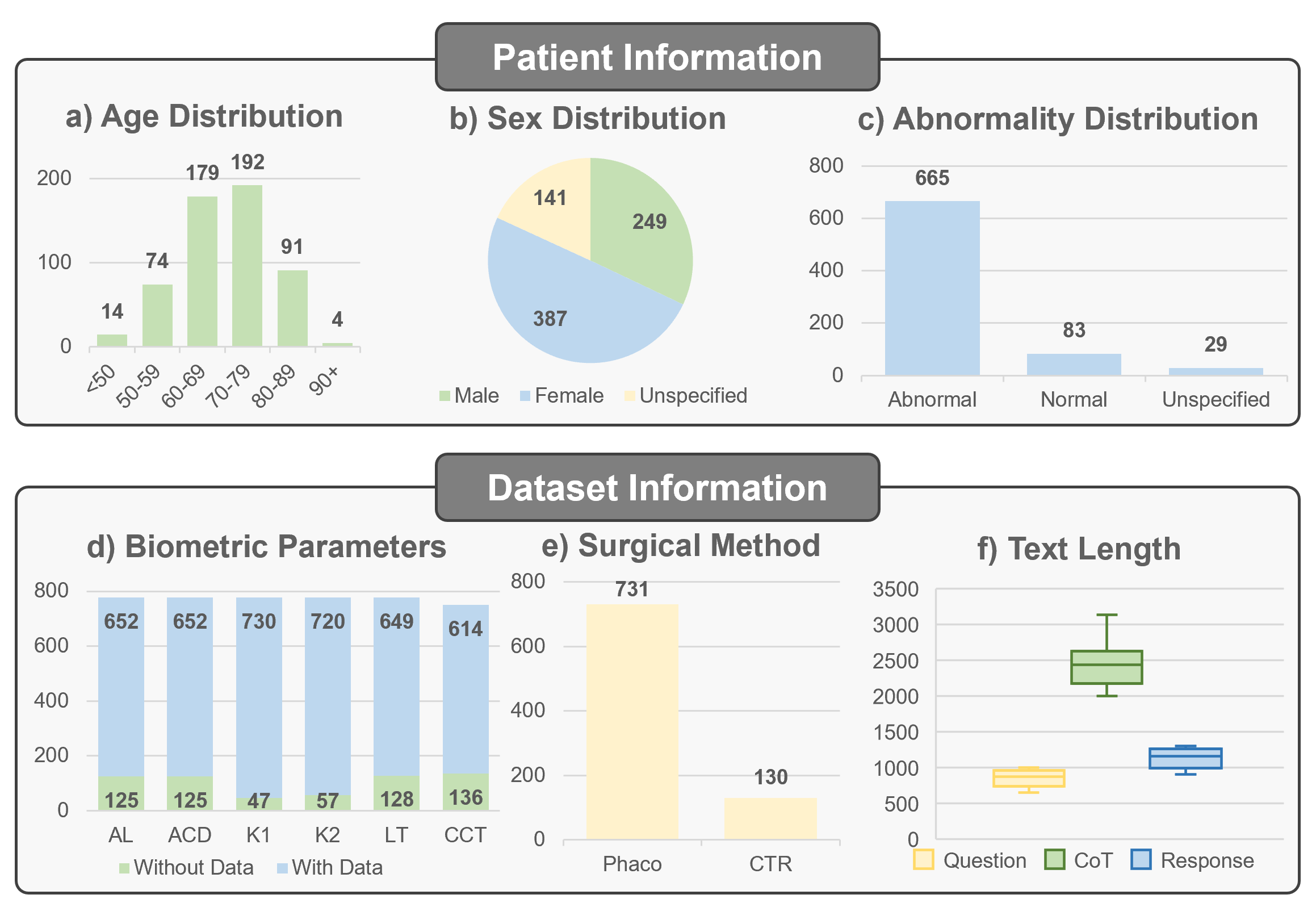}
    \vspace{-2ex}
    \caption{Summary statistics of the cataract surgery dataset used for developing and evaluating LLMs. The upper panel displays: a) Age Distribution (mean: 69.4 years, median: 70.0 years, SD: 10.5 years), b) Sex Distribution, c) Abnormality Distribution. The bottom panel displays: d) Biometric Parameter Availability, e) Recommended Surgical Method (Phaco means Phacoemulsification), and f) Text Length Statistics for Question, CoT, and Response fields. Unspecified categories refer to missing or unreported data.}
    \vspace{-1ex}
    \label{fig:data_distr}
\end{figure*}

%While CataractSurg-80K is a large-scale dataset we constructed, we selected a benchmark subset of 777 cases from it—sourced from the electronic medical records of Peking University Third Hospital—for model performance evaluation.
%Different from CataractSurg-80K, we constructed a benchmark dataset of 777 cases from the electronic medical records of Peking University Third Hospital to evaluate the performance of different models. This dataset covers all common cataract types and contains a small number of normal (non-cataract) cases as negative controls. All data were de-identified and ethically approved.
CataractSurg-80K is a large-scale dataset curated from ophthalmology electronic medical records at Peking University Third Hospital. For our evaluation, we chose a benchmark subset of 777 cases instead of using the entire dataset. This subset includes all common types of cataracts, along with a small number of normal (non-cataract) cases to serve as negative controls. All data were de-identified and received ethical approval.

Figure~\ref{fig:data_distr} summarizes the main features of the benchmark dataset. The age distribution indicates that the patients seeking medical treatment are primarily elderly, with an average age of 69.4 years. This aligns with the typical cataract population, which also has a high percentage of women. Most cases in the dataset are classified as abnormal, while a small number are marked as normal or unspecified to simulate the reality of missing data in clinical settings.

We also analyzed the availability of six key biometric parameters: AL, ACD, K1, K2, LT, and CCT. Most patients had complete measurement data, while some records were missing specific values due to incomplete documentation or inaccurate data extraction, and certain cases lacked essential information. The majority of patients underwent phacoemulsification, while a small number opted for capsular tension ring (CTR) implantation. Finally, we provided text length statistics for three categories in the dataset: question, chain of thought, and response, to offer a clearer understanding of the complexity and depth of annotation within the dataset.

\subsection{Benchmark Experiment}
We conducted quantitative evaluations on general LLMs and customized fine-tuned models for this work on two core tasks: chain of thought generation (Question -> CoT) and final surgical advice (Question -> Response). The evaluation indicators cover text generation quality (BLEU, ROUGE-L, BERTScore), key term extraction (k-F1, k-Precision, k-Recall), entity-level accuracy (Entity F1/Precision/Recall), and chain of reasoning consistency (NLI Consistency), which comprehensively reflect the performance of the model in multiple dimensions such as medical facts, clinical terms, reasoning continuity, and logical consistency.

For text generation quality, we selected three mainstream automatic evaluation indicators: BLEU, ROUGE-L, and BERTScore. BLEU mainly examines the precise phrase matching at the n-gram level, emphasizing the accurate restoration of medical terms and professional expressions; ROUGE-L focuses on the longest common subsequence, which is more suitable for evaluating the overall structure of the reasoning chain, logical integrity and key point coverage; BERTScore is based on deep semantic vectors, which can evaluate the similarity between the generated text and the reference answer in the semantic space, and has stronger adaptability to diversified medical expressions. The experimental results show that after domain fine-tuning, the model significantly outperforms the general large model across all three metrics. The most notable improvement is in BLEU, highlighting that the injection of medical knowledge enhances the accurate restoration of professional terminology. Although the BERTScore baseline is already high, the improvements are modest, indicating that mainstream large models excel in general semantic understanding.

By reviewing Table \ref{tab:benchmark_text}, we find that our fine-tuned model Qwen-CSP outperforms mainstream general models across all three indicators. BLEU is sensitive to exact n-gram matching, which means our Qwen-CSP shows significant improvement when domain knowledge is included, especially in restoring specialized terms. BERTScore, which assesses deep semantic similarity, has a higher tolerance for diverse expressions, but improvements are limited for models already strong in general semantics. In general, our method excels in the restoration of professional terminology and structured medical expressions while maintaining a strong advantage in semantic understanding.
\begin{table}[ht]
\centering
\scriptsize
\caption{Performance of Baseline Models on the Cataract Surgery Benchmark Dataset (BLEU, ROUGE, BERTScore, Keypoint-level Metrics)}
\label{tab:benchmark_text}
\begin{tabular}{lccccccc}
\toprule
\textbf{Model} & \textbf{Task} & \textbf{BLEU} & \textbf{ROUGE-L} & \textbf{BERTScore\textsubscript{F1}} & \textbf{k-F1} & \textbf{k-Prec.} & \textbf{k-Rec.} \\
\midrule
GPT4.1 & COT     & 0.039$\pm$0.018 & 0.197$\pm$0.029 & 0.865$\pm$0.013 & \underline{0.225$\pm$0.038} & 0.230$\pm$0.039 & \underline{0.230$\pm$0.051} \\
GPT4.1 & Response &0.068$\pm$0.025 & 0.224$\pm$0.040 & 0.867$\pm$0.014 & 0.259$\pm$0.050 & 0.224$\pm$0.038 & \underline{0.325$\pm$0.078} \\
Claude-3.7 & COT & 0.034$\pm$0.014 & 0.186$\pm$0.032 & 0.853$\pm$0.014 & 0.204$\pm$0.042 & 0.225$\pm$0.042 & 0.197$\pm$0.050 \\
Claude-3.7 & Response & 0.040$\pm$0.018 & 0.206$\pm$0.038 & 0.854$\pm$0.016 & 0.231$\pm$0.054 & 0.243$\pm$0.041 & 0.226$\pm$0.063 \\
Gemini-2 & COT & 0.023$\pm$0.012 & 0.160$\pm$0.028 & 0.846$\pm$0.017 & 0.196$\pm$0.042 & 0.197$\pm$0.042 & 0.201$\pm$0.051 \\
Gemini-2 & Response & 0.020$\pm$0.022 & 0.141$\pm$0.064 & 0.830$\pm$0.021 & 0.169$\pm$0.096 & 0.216$\pm$0.077 & 0.194$\pm$0.141 \\
DeepSeek & COT & 0.054$\pm$0.018 & \underline{0.212$\pm$0.023} & \underline{0.876$\pm$0.011} & 0.207$\pm$0.030 & 0.224$\pm$0.034 & 0.194$\pm$0.033 \\
DeepSeek & Response & 0.019$\pm$0.013 & 0.182$\pm$0.025 & 0.856$\pm$0.010 & 0.170$\pm$0.034 & 0.162$\pm$0.035 & 0.182$\pm$0.039 \\
Qwen-Max & COT & \underline{0.059$\pm$0.017} & 0.207$\pm$0.024 & 0.863$\pm$0.012 & 0.220$\pm$0.033 & \underline{0.256$\pm$0.043} & 0.196$\pm$0.034 \\
Qwen-Max & Response & \underline{0.098$\pm$0.038} & \underline{0.249$\pm$0.049} & \underline{0.873$\pm$0.015} & \underline{0.270$\pm$0.054} & \underline{0.246$\pm$0.050} & 0.312$\pm$0.074 \\
Qwen-8B & COT & 0.047$\pm$0.015 & 0.153$\pm$0.015 & 0.831$\pm$0.011 & 0.214$\pm$0.031 & 0.219$\pm$0.034 & 0.212$\pm$0.034 \\
Qwen-8B & Response & 0.041$\pm$0.012 & 0.173$\pm$0.020 & 0.837$\pm$0.012 & 0.218$\pm$0.037 & 0.170$\pm$0.028 & 0.311$\pm$0.066 \\
Qwen-4B & COT & 0.048$\pm$0.016 & 0.153$\pm$0.016 & 0.829$\pm$0.013 & 0.216$\pm$0.030 & 0.230$\pm$0.034 & 0.205$\pm$0.031 \\
Qwen-4B & Response & 0.039$\pm$0.012 & 0.166$\pm$0.023 & 0.836$\pm$0.013 & 0.222$\pm$0.039 & 0.179$\pm$0.030 & 0.302$\pm$0.064 \\
Qwen-CSP & COT & \textbf{0.131$\pm$0.029} & \textbf{0.261$\pm$0.034} & \textbf{0.886$\pm$0.015} & \textbf{0.292$\pm$0.043} & \textbf{0.317$\pm$0.046} & \textbf{0.275$\pm$0.045} \\
Qwen-CSP & Response & \textbf{0.170$\pm$0.041} & \textbf{0.330$\pm$0.044} & \textbf{0.896$\pm$0.013} & \textbf{0.367$\pm$0.054} & \textbf{0.377$\pm$0.062} & \textbf{0.361$\pm$0.051} \\
\midrule
\textit{Improvement} & COT & 122\% & 23.1\% & 1.14\% & 15.8\% & 23.8\% & 19.6\% \\
\textit{Improvement} & Response & 73.4\% & 32.5\% & 2.63\% & 35.9\% & 53.3\% & 11.1\% \\
\bottomrule
\end{tabular}
\captionsetup{font=footnotesize}
\caption*{This table shows the performance of each model on text generation and keypoint metrics, including BLEU, ROUGE-L, BERTScore\textsubscript{F1}, and keypoint-level F1, precision, and recall ($k$-F1, $k$-Prec., $k$-Rec.). Bold indicates the best result; underline indicates the second-best.}
\end{table}

To further evaluate the model's ability to extract key medical information, including core terms and entities in the medical field, covering surgical methods, anatomical structures, disease names, etc., Scispacy entity recognition and automatic extraction of medical terminology vocabulary were combined. We designed three indicators: k-F1, k-Precision, and k-Recall. The key information k-Precision represents the proportion of correctly restored terms in the generated text, k-Recall measures the coverage of key information in the reference answer, and k-F1 is the harmonic average of the two. Table~\ref{tab:benchmark_structure}
has shown our Qwen-CSP achieved the best results in all three indicators, especially in k-Recall and k-F1, which showed that domain adaptation not only improved the accuracy of medical entity recognition but also enhanced the coverage of key information, while effectively suppressing the generation of irrelevant or illusory entities.
\begin{table}[ht]
\centering
\caption{Performance of Baseline Models on the Cataract Surgery Benchmark Dataset (Entity-level and NLI Consistency)}
\label{tab:benchmark_structure}
\footnotesize 
\begin{tabular}{lccccc}
\toprule
\textbf{Model} & \textbf{Task} & \textbf{Entity F1} & \textbf{Entity Prec.} & \textbf{Entity Rec.} & \textbf{NLI Cons.} \\
\midrule
GPT4.1 & COT & \underline{0.247$\pm$0.058} & 0.244$\pm$0.067 & \textbf{0.279$\pm$0.071} & \underline{0.907$\pm$0.276} \\
Claude-3.7 & COT & 0.208$\pm$0.062 & \underline{0.246$\pm$0.076} & 0.215$\pm$0.078 & 0.619$\pm$0.466 \\
Gemini-2 & COT & 0.214$\pm$0.059 & 0.211$\pm$0.064 & 0.252$\pm$0.077 & \textbf{0.916$\pm$0.264} \\
DeepSeek & COT & 0.201$\pm$0.065 & 0.245$\pm$0.077 & 0.204$\pm$0.069 & 0.201$\pm$0.362 \\
Qwen-Max & COT & 0.096$\pm$0.079 & 0.099$\pm$0.098 & 0.167$\pm$0.106 & 0.297$\pm$0.321 \\
Qwen-8B & COT & 0.221$\pm$0.057 & 0.233$\pm$0.063 & 0.248$\pm$0.080 & 0.823$\pm$0.350 \\
Qwen-4B & COT & 0.226$\pm$0.056 & 0.243$\pm$0.064 & 0.245$\pm$0.076 & 0.859$\pm$0.328 \\
Qwen-CSP & COT & \textbf{0.258$\pm$0.083} & \textbf{0.294$\pm$0.096} & \underline{0.270$\pm$0.082} & 0.158$\pm$0.324 \\
\midrule
\textit{Improvement} & COT & 4.45\% & 19.5\% & -3.3\% & -- \\
\bottomrule
\end{tabular}
\captionsetup{font=footnotesize}
\caption*{This table presents structured output evaluation, including entity-level F1/precision/recall and NLI consistency. Missing metrics are marked as ``--''. Bold indicates best value; underline indicates second-best.}
\end{table}

For the CoT generation task, we further introduce stepwise entity-level evaluation and NLI-based consistency metrics. The entity-level F1, Precision, and Recall are computed by automatically segmenting each reasoning step, extracting medical entities, and comparing them with the reference at each step, thereby quantifying the factual alignment of the reasoning chain. NLI consistency leverages a pre-trained natural language inference model to assess the logical relationship between each pair of consecutive reasoning steps, reporting the proportion of “entailment” as an indicator of overall logical coherence.

The experimental results are shown in Table~\ref{tab:benchmark_structure}. The fine-tuned model performs outstandingly in the step-by-step medical entity recognition task, with leading F1 value, precision and recall. However, the NLI consistency score dropped from 0.859 to 0.158 after domain fine-tuning, indicating that although the model's ability to restore medical details has improved, the logical coherence of the reasoning chain has weakened. The reason is that fine-tuning prompts the model to generate richer and more diverse clinical content, which increases the amount of information while also increasing logical fluctuations. To address this problem, in the future, step-by-step NLI discrimination or medical knowledge graph constraints can be introduced in training to take into account both the richness of medical expressions and the consistency of the reasoning chain.

\subsection{Ablation study}
To rigorously evaluate the effectiveness of the main components of our training pipeline, we performed a series of ablation experiments on the CataractSurg-80K dataset and evaluated on the benchmark. The transition from a base model to a clinical model capability is very complex, as shown in the figure \ref{fig:Intro}. We used a multi-stage training approach: First, we performed supervised fine-tuning (SFT) on a large-scale cataract dataset to enhance the model's domain knowledge \textbf{(A)}. Subsequently, we performed two-stage task-specific training using our carefully selected dataset, first generating a CoT reasoning process \textbf{(B)} and then generating the final surgical recommendations \textbf{(C)}. Ablation experiments aim to quantify the respective contributions of domain knowledge adaptation and CoT supervision in this pipeline.

\begin{table}[ht]
\centering
\scriptsize
\caption{Ablation results on the Cataract Surgery Benchmark (BLEU, ROUGE, BERTScore, Keypoint-level Metrics. The term "Base" refers to the base model Qwen3-4B.}
\label{tab:ablation_text}
\begin{tabular}{lccccccc}
\toprule
\textbf{Model} & \textbf{Task} & \textbf{BLEU} & \textbf{ROUGE-L} & \textbf{BERTScore\textsubscript{F1}} & \textbf{k-F1} & \textbf{k-Prec.} & \textbf{k-Rec.} \\
\midrule
Base & COT & 0.048$\pm$0.016 & 0.153$\pm$0.016 & 0.829$\pm$0.013 & 0.216$\pm$0.030 & 0.230$\pm$0.034 & 0.205$\pm$0.031 \\
Base & Response & 0.039$\pm$0.012 & 0.166$\pm$0.023 & 0.836$\pm$0.013 & 0.222$\pm$0.039 & 0.179$\pm$0.030 & 0.302$\pm$0.064 \\
Base+B+C & COT & 0.103$\pm$0.025 & 0.193$\pm$0.022 & 0.845$\pm$0.011 & 0.289$\pm$0.036 & 0.297$\pm$0.042 & \textbf{0.283$\pm$0.037} \\
Base+B+C & Response & 0.102$\pm$0.030 & 0.231$\pm$0.028 & 0.855$\pm$0.013 & 0.350$\pm$0.055 & 0.313$\pm$0.061 & \textbf{0.404$\pm$0.053} \\
Base+A+C & COT & 0.056$\pm$0.016 & 0.226$\pm$0.025 & 0.861$\pm$0.011 & 0.236$\pm$0.035 & \textbf{0.373$\pm$0.057} & 0.173$\pm$0.028 \\
Base+A+C & Response & 0.169$\pm$0.040 & \textbf{0.335$\pm$0.044} & 0.894$\pm$0.013 & \textbf{0.375$\pm$0.056} & 0.393$\pm$0.066 & 0.362$\pm$0.050 \\
Base+A+B+C & COT & \textbf{0.131$\pm$0.029} & \textbf{0.261$\pm$0.034} & \textbf{0.886$\pm$0.015} & \textbf{0.292$\pm$0.043} & 0.317$\pm$0.046 & 0.275$\pm$0.045 \\
Base+A+B+C & Response & \textbf{0.170$\pm$0.041} & 0.330$\pm$0.044 & \textbf{0.896$\pm$0.013} & 0.367$\pm$0.054 & \textbf{0.377$\pm$0.062} & 0.361$\pm$0.051 \\
\bottomrule
\end{tabular}
%\caption*{Note that here we refer to Qwen3-4B as the base model named "Base"}
\end{table}

\begin{table}[ht]
\centering
\caption{Ablation results on the Cataract Surgery Benchmark (Entity-level and NLI Consistency)}
\label{tab:ablation_structured}
\footnotesize
\begin{tabular}{lccccc}
\toprule
\textbf{Model} & \textbf{Task} & \textbf{Entity F1} & \textbf{Entity Prec.} & \textbf{Entity Rec.} & \textbf{NLI Cons.} \\
\midrule
Base & COT & 0.226$\pm$0.056 & 0.243$\pm$0.064 & 0.245$\pm$0.076 & 0.859$\pm$0.328 \\
Base+B+C & COT & \textbf{0.293$\pm$0.079} & 0.299$\pm$0.085 & \textbf{0.324$\pm$0.088} & 0.829$\pm$0.365 \\
Base+A+C & COT & 0.279$\pm$0.057 & \textbf{0.396$\pm$0.097} & 0.232$\pm$0.060 & \textbf{0.976$\pm$0.141} \\
Base+A+B+C & COT & 0.258$\pm$0.083 & 0.294$\pm$0.096 & 0.270$\pm$0.082 & 0.158$\pm$0.324 \\
\bottomrule
\end{tabular}
\end{table}

We first ablated the initial supervised fine-tuning on the cataract dataset and trained the model directly on the task dataset without prior domain adaptation. As shown in Table~\ref{tab:ablation_text} and Table~\ref{tab:ablation_structured}, the removal of prior domain adaptation led to a consistent performance drop in all evaluation metrics for CoT and response tasks. In particular, clinical reliability and accurate extraction of key medical points (k-F1, k-Precision, k-Recall) were adversely affected, highlighting the critical role of underlying domain knowledge in enabling the model to reliably capture and reproduce medical concepts and entities.

We secondly removed explicit CoT annotations during training so that the model is only supervised to generate final recommendations without intermediate reasoning steps. As shown in Table~\ref{tab:ablation_text} and Table~\ref{tab:ablation_structured}, the absence of CoT supervision led to a further drop in interpretability and overall performance, as evidenced by the drop in BLEU, ROUGE-L, BERTScore, and entity-level metrics. This highlights the importance of incorporating CoT annotations in training to guide the model toward more explainable and clinically grounded decision-making processes.

Interestingly, the partial ablation model outperformed the full model on some secondary indicators (such as entity-level F1 and NLI Consistency). We believe this is because after domain adaptation and reasoning chain supervision, the model output has become more detailed and diverse, which improves clinical interpretability but may lead to certain variations in entity expression and reasoning steps. The ablation model tends to be more templated and simplifies output, so it performs better in consistency and entity overlap, but the amount of information and reference value in actual clinical decision-making are limited.
\subsection{Discussion} \label{discussion}

The results of this study fully verify the effectiveness of structured multi-agent processes in AI-assisted preoperative decision-making for cataract surgery. By combining multi-agent clinical information understanding, enhanced reasoning supervision, and domain knowledge fine-tuning, our framework significantly outperforms general and existing large medical models in multiple clinical tasks such as information extraction, IOL recommendation, and surgical feasibility assessment, and performs well in interpretability and reasoning transparency. This not only improves the credibility of the model in actual clinical applications but also provides a new paradigm for AI assistance in complex medical decision-making tasks.

It is worth noting that the multi-agent structure and multi-stage training strategy can effectively cope with the complexity of heterogeneous multi-source clinical data and realize the automation of the entire process from original reports to structured reasoning and final recommendations. In the experiment, ablation analysis further confirmed the key role of basic domain knowledge guidance and CoT supervision in improving the clinical performance and reasoning quality of the model. In addition, the support of large-scale, high-quality datasets enables the model to have stronger generalization ability and stability. These achievements provide a solid foundation for the development of intelligent decision-making systems in ophthalmology and the wider medical field.

\section{Limitations}\label{limitation}
We acknowledge that there are still some limitations in the study. First, our architecture is capable of handling multiple reports, but due to funding constraints, this study focuses on four key ophthalmic reports that are the most influential and selected by field experts. This simplification may sometimes miss key information for physician decision-making.

Second, the current model size is relatively small and may be limited in its ability to capture the full complexity of rare or atypical cases. The increase in the number of rare diseases in the dataset and the number of model parameters will address these issues.

The current evaluation lacks validation in clinical practice. To address this, we have developed a user interface for a pilot deployment that applies the model to clinical scenarios, aiming to reduce the costs associated with doctor-patient communication. However, further research involving direct interactions with clinicians and patients is still ongoing.

\section{Conclusion}
% 局限性：缺少临床实践的结果，我们已经设计了UI界面尝试让医生和患者使用该模型进行实践
% 局限性：只考虑了4种报告，虽然是医生筛选的最重要的四个报告，但是根据之前的分析，每个患者情况不一致，检查结果也不一样，我们对真实的术前准备和报告分析做了简化
% 局限性：模型的局限性-模型体量小-能力。
% 结论一：我们提出了数据清洗pipeline，专注信息提取而非分析数据，比较客观的清洗大量复杂结构数据，拓展更多场景。结论二：首次提出白内障手术术前诊断数据集和benchmark，医疗大模型不再局限于考试和问答上，在更复杂的应用场景下进行了。结论三：混合方法微调qwen3，端到端的解读和手术规划。
In this study, we proposed a comprehensive AI-driven preoperative decision support framework for cataract surgery. First, we designed a novel multi-agent reasoning process and multi-stage fine-tuning strategy to achieve structured understanding and reasoning of complex clinical information, effectively improving the intelligence level of preoperative decision-making. Second, we built a large-scale, expert-annotated CataractSurg-80K dataset, and finely annotated 777 benchmarks based on real cases, systematically supporting the training and objective evaluation of the model. We also open-sourced a set of multi-stage large model training methods, which achieved an organic connection from step-by-step clinical reasoning to end-to-end surgical planning, and improved the interpretability and practical application value of the model.
%Looking forward, we hope that this method and resources will accelerate the development of safe, scalable, and explainable medical AI systems. By promoting open source collaboration based on privacy-protected real-world clinical data, we look forward to promoting technological progress in ophthalmology and other medical fields that require highly reliable and explainable AI solutions. 
Our datasets and benchmarks provide a foundation for industry-standard evaluation and model reproducibility research. Our open-source training framework also provides a powerful tool for the community to conduct large-scale clinical reasoning and decision support research.

\newpage
\bibliographystyle{unsrtnat}
\bibliography{references}
%%%%%%%%%%%%%%%%%%%%%%%%%%%%%%%%%%%%%%%%%%%%%%%%%%%%%%%%%%%%

\newpage
\appendix

% \section{Technical Appendices and Supplementary Material}
% Technical appendices with additional results, figures, graphs and proofs may be submitted with the paper submission before the full submission deadline (see above), or as a separate PDF in the ZIP file below before the supplementary material deadline. There is no page limit for the technical appendices.
\section{Data}
\subsection{Open-sourced Multi-source Cataract Medical QA/CoT Datasets}

\textbf{License.} All resources are released under the \textbf{CC BY-NC-SA 4.0} license.

\textbf{Training.} We provide a collection of Chinese and English datasets for supervised fine-tuning and evaluation in the field of cataract clinical reasoning and report understanding.

\begin{itemize}
    \item \textbf{Base Knowledge} \\
    \url{https://huggingface.co/datasets/my2000cup/CataractSurg-80K} \\
    source PDF file comes from: \\
    \url{https://apacrs.org/ppp_cataract_EN/files/assets/common/downloads/APACRS-PPP-Cataract%20Surgery..pdf} \\
    \url{https://apacrs.org/ppp_cataract_cn/files/assets/common/downloads/APACRS-PPP-Cataract%20Surgery-cn.pdf}
    \item \textbf{CataractSurg-80K:} \\
    \url{https://huggingface.co/datasets/my2000cup/CataractSurg-80K} \\
    cataract\_base\_zh, cataract\_base\_en: Instruction-based QA (Chinese/English) \\
    inference01\_zh, inference01\_en: Chain-of-Thought reasoning \\
    inference02\_zh, inference02\_en: Clinical recommendations \\
    All subsets follow the Alpaca SFT format.
\end{itemize}

\textbf{Model.} 

\begin{itemize}
    \item \textbf{Qwen-CSP:} \\
    \url{https://huggingface.co/my2000cup/Qwen-CSP} \\
    Fine-tuned Qwen3-4B model on the CataractSurg-80K datasets.
\end{itemize}

\textbf{Code.} 

\begin{itemize}
    \item \textbf{CataractCareMAS:} \\
    \url{https://anonymous.4open.science/r/CataractCareMAS-A363} \\
    Data processing, model training, and evaluation code.
\end{itemize}

\subsection{Notes on open source datasets and code}
Due to the requirements of patient privacy protection and ethical compliance, we cannot open-source the original patient electronic medical record data, like the original data before preprocessing by the multi-agent system. Therefore, the multi-agent cleaning and structuring process for the original data cannot be directly reproduced by a third party. The actual open source dataset only contains the downstream training data that has been strictly desensitized and structured. This part of the data has removed all personal identity information and obtained hospital ethics approval.

Although the complete pipeline cannot be directly opened to the public, we still open source the code of multi-agent data cleaning and structuring. The code has good versatility and scalability and can be migrated to other medical or multimodal scenarios. Researchers can combine their own legal and compliant data resources and quickly implement data structuring and cleaning processes suitable for their own tasks based on our code framework. For different application scenarios, users need to make appropriate adjustments and optimizations to the role division, rules, and prompt word engineering of multi-agents based on actual needs.

We believe that with the opening of more compliant data, the multi-agent data cleaning framework will provide valuable reference and tool support for medical AI data engineering, structured knowledge extraction and other fields.

\section{Prompts}
%\newpage
\subsection{Prompts for Heterogeneous Ophthalmic Reports Interpretation}
\begin{figure}
    \centering
    \includegraphics[width=1.0\textwidth]{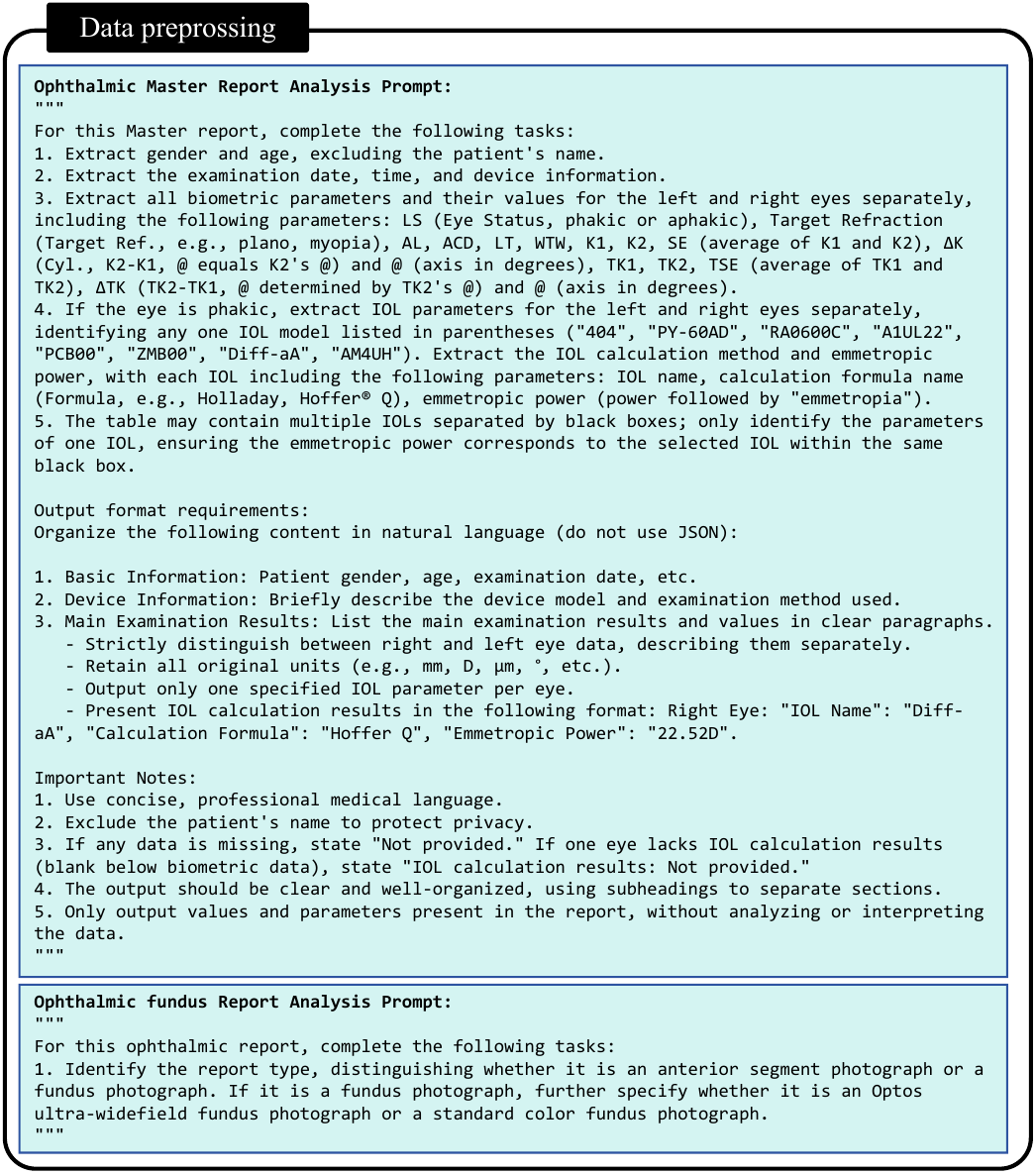}
    \caption{Prompt template for interpreting ophthalmic master report and fundus report.}
    \label{train}
\end{figure}

\begin{figure}
    \centering
    \includegraphics[width=1.0\textwidth]{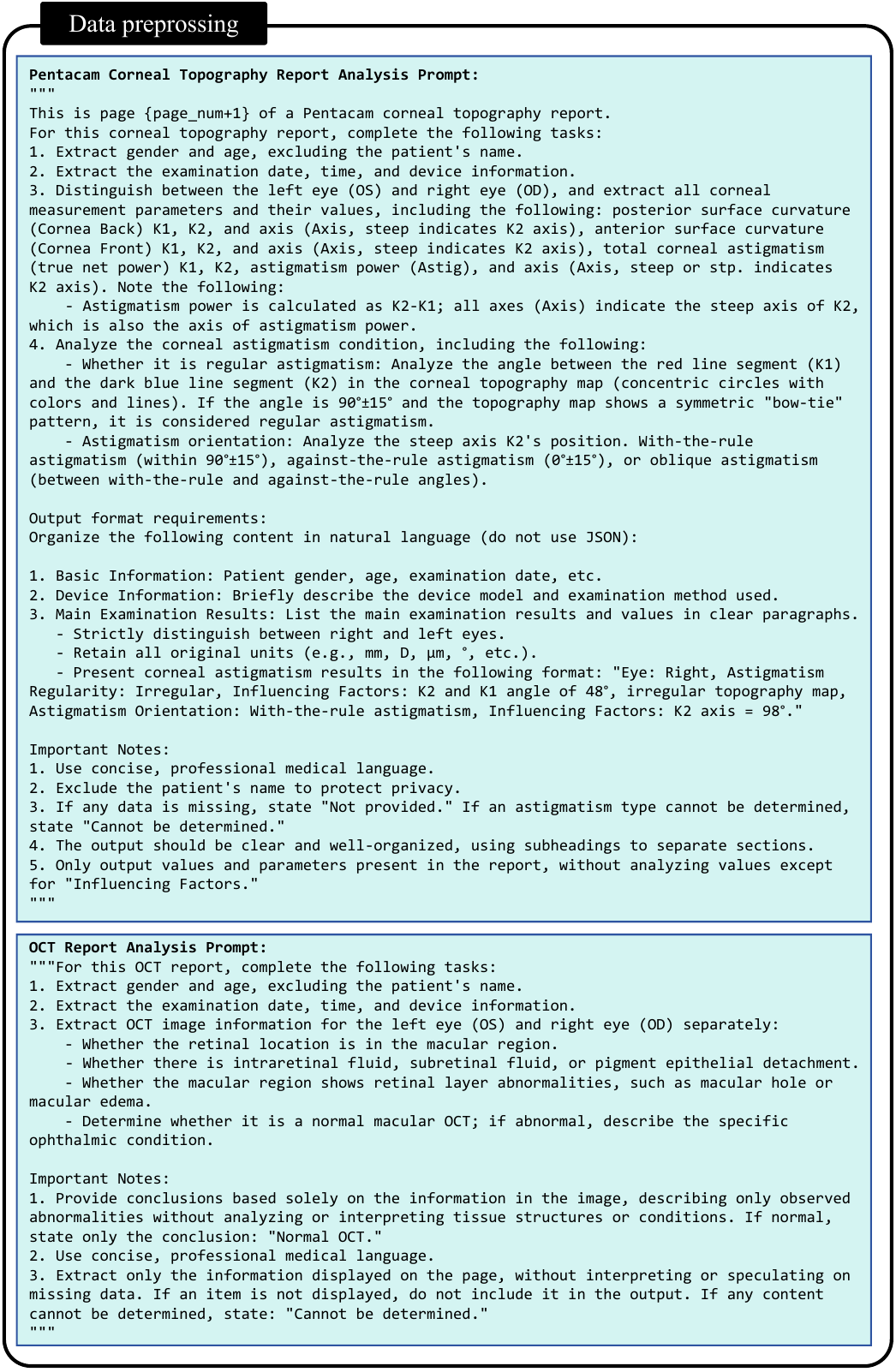}
    \caption{Prompt template for interpreting Pentacam Corneal Topography and OCT reports.}
    \label{train1}
\end{figure}

\newpage
\subsection{Prompts for Constructing Reasoning-enhanced Datasets}

\begin{figure}[htbp!]
\vspace{-1ex}
    \centering
    \includegraphics[width=1\textwidth]{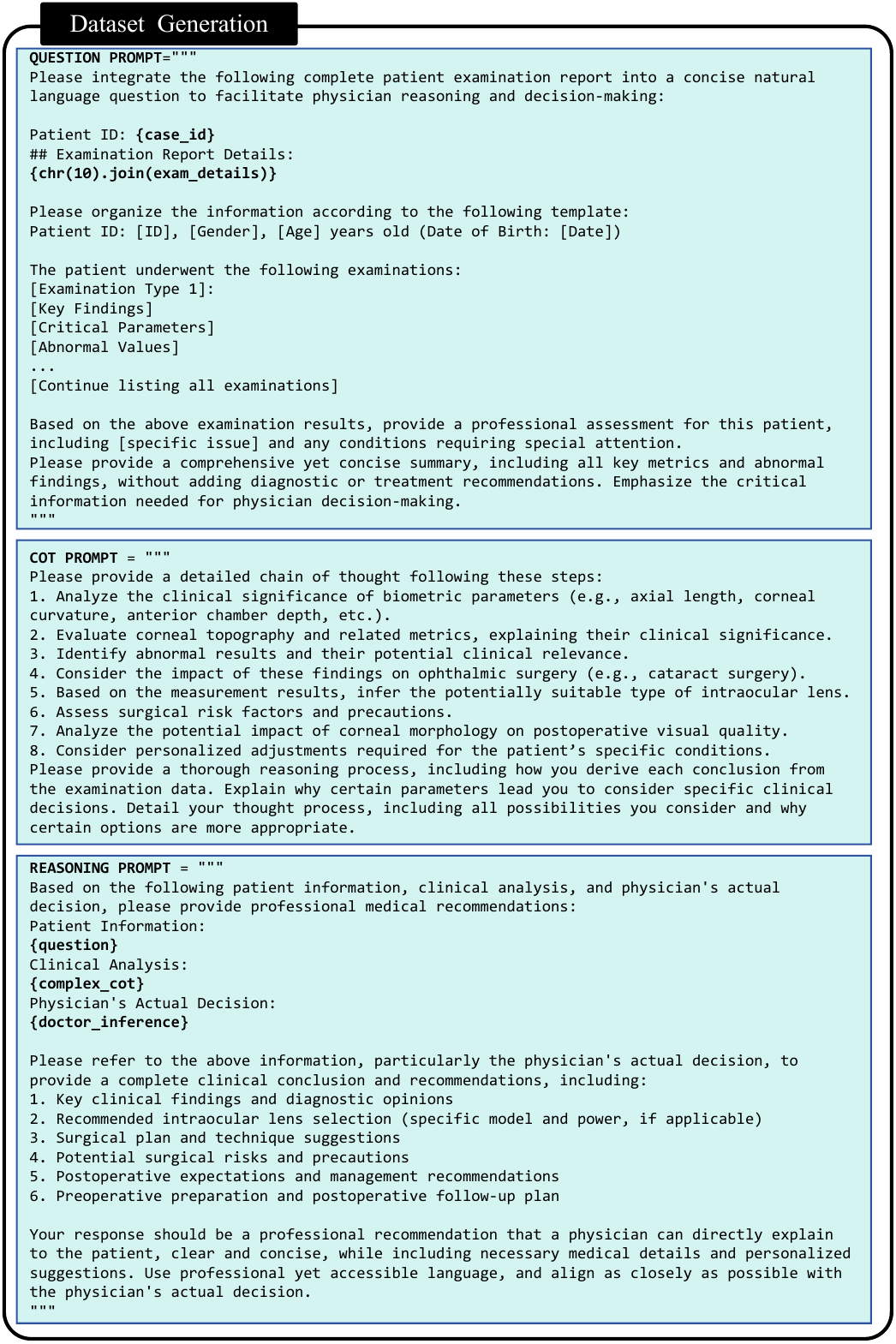}
    \vspace{-1ex}
    \caption{Prompt designed and utilized for constructing reasoning-enhanced datasets.}
    \label{train2}
\end{figure}

\section{Case Study}
\begin{figure}[htbp!]
\vspace{-1ex}
    \centering
    \includegraphics[width=1\textwidth]{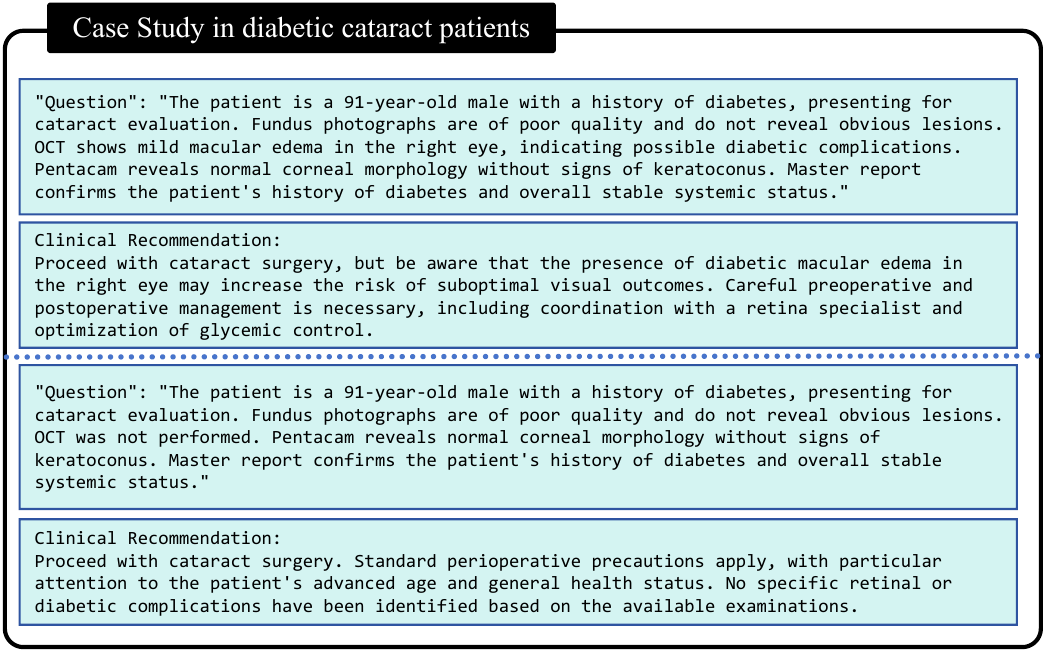}
    \vspace{-1ex}
    \caption{Impact of OCT Missing on Clinical Reasoning of Multi-agent Systems}
    \label{case0}
\end{figure}

We chose a representative case in cataract surgery, namely, surgical risk reminders for diabetic patients. In this case, the OCT report detected diabetes, so the relevant reminder appeared in our Question column. We simulated the lack of OCT and observed significant differences in clinical recommendations for patients with diabetic cataracts.

When the OCT report was available, the multi-agent system was able to identify diabetic macular edema, a key complication that directly affects surgical prognosis and postoperative management. In contrast, when the OCT report was missing, the system failed to identify this key risk, resulting in generic recommendations that may overlook important diabetic complications. The result summary is shown in Figure \ref{case0}

This case highlights the advantage of our multi-agent approach: the more comprehensive the information obtained and processed, the more reliable the clinical reasoning. Integrating a full set of reports enables the system to provide clinicians with more accurate, personalized, and safer decision support.

\begin{figure}[htbp!]
\vspace{-1ex}
    \centering
    \includegraphics[width=1\textwidth]{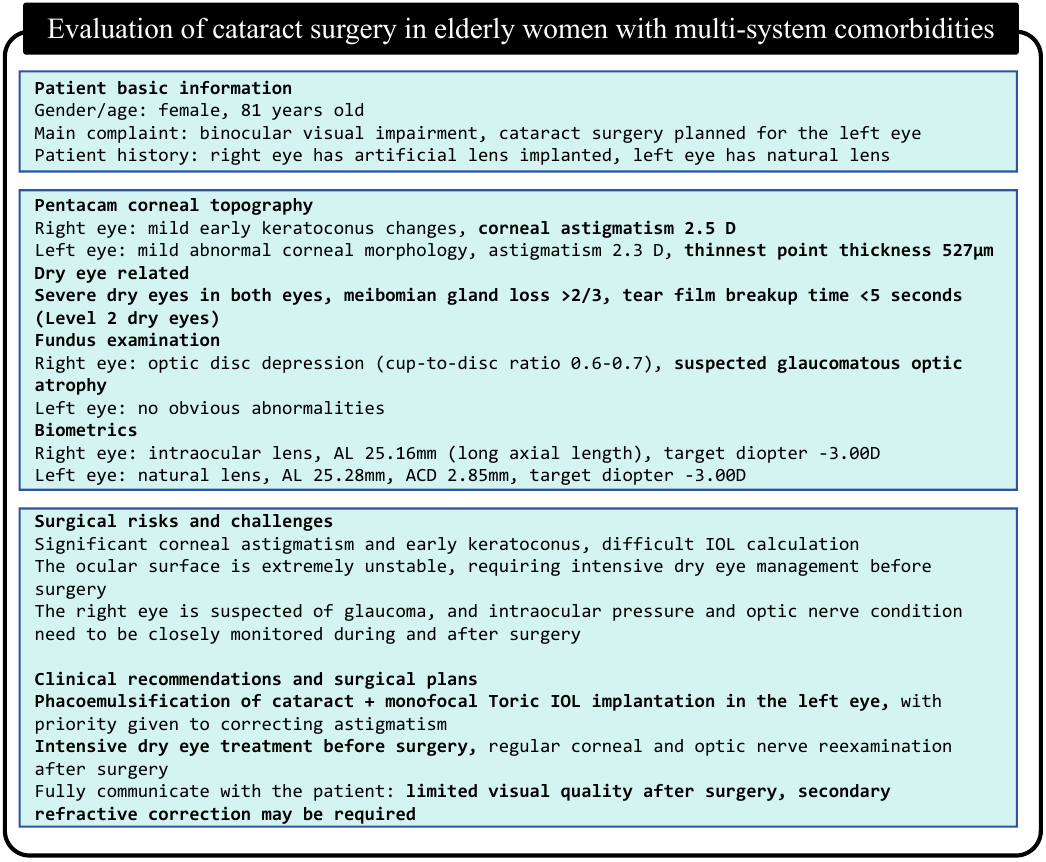}
    \vspace{-1ex}
    \caption{Impact of OCT Missing on Clinical Reasoning of Multi-agent Systems}
    \label{case1}
\end{figure}

We also selected an elderly female patient with multiple complications as a complex case for testing. The patient also had risk factors such as abnormal corneal morphology, significant astigmatism, dry eyes, and glaucoma, which placed high demands on the preoperative evaluation and surgical selection of cataract surgery. The experimental results showed that the multi-agent system can effectively integrate multi-source examination data, accurately identify the main risk points of patients, and give surgical recommendations that are highly consistent with the opinions of real clinical experts. This result further verifies that our method has good reliability and clinical practical value in actual clinical scenarios with complex and multi-factor interventions. Result summary shown in Figure \ref{case1}.

\section{Further Discussion}
\subsection{Discussion of social impact, ethics, and limitations}
Our system is expected to improve the accessibility and fairness of healthcare by empowering clinicians in resource-poor areas. Multi-agent systems can improve doctors' efficiency in diagnosing common diseases, especially in primary care institutions, and help alleviate the problem of uneven distribution of high-quality medical resources. Most diseases encountered in clinical scenarios are routine cases, so this has practical significance for improving the overall level of medical care. However, in the face of complex or rare diseases, artificial intelligence systems still have a long way to go and need to continue to accumulate data and optimize reasoning capabilities.

From the perspective of multimodal and multi-source data processing, multi-agent systems have stronger professionalism and scalability than single, general large models. This is also the original intention of our development and introduction of multi-agent collaboration to process different types of medical reports. By focusing on different subsystems on different medical subtasks, the processing efficiency and professional reliability of the output can be effectively improved.

In terms of ethics and privacy, the bottleneck of the promotion of medical large models is often not the technology itself, but the high attention to data privacy security and ethical compliance. Multi-agent systems have the natural advantages of data classification and decoupling, which help to better achieve de-identification and privacy protection during data flow and processing. This study strictly desensitized and standardized the original data under the hospital's ethical approval and compliance license, providing a model for the safe application of subsequent artificial intelligence systems.

It is worth emphasizing that our tool only serves as a clinical decision support system to assist doctors in diagnosis and treatment judgment, rather than replacing doctors for autonomous diagnosis or treatment. We recommend that manual review and final review should always be maintained during use to prevent new risks caused by technology abuse or over-reliance on automated suggestions. At the same time, it is recommended that the mechanism construction in ethical governance, data security, and model interpretability should be continuously strengthened in the subsequent actual application and promotion process.

\subsection{Preparation and impact discussion of basic knowledge SFT dataset}
Preparation and impact discussion of basic knowledge SFT dataset. In the first stage of training our large model, the preparation of the basic knowledge dataset also adopted a multi-agent system, but its structure was much simpler than that of the cataract surgery multi-agent system, focusing mainly on the automatic organization of medical knowledge and question-answer generation. The specific process is as follows:

First, the data comes from the internationally recognized authoritative guide-APACRS "Patient Guide for Cataract Surgery" (Chinese and English versions, link in the appendix). We removed the references and content before the table of contents in the original file, and only retained the pure text information. For the convenience of subsequent processing, the text content is divided into several blocks of about 5,000 words each.

For each small block, GPT4o-mini is used as the core of the agent, and professional questions related to the text content are automatically generated through the prompt word engineering, and further answers based on the text block are generated. In order to improve the accuracy and authority of the answers, the system also calls the ReAct Agent to automatically retrieve relevant Wikipedia content to supplement and unify the answers. Through the above process, a high-quality SFT paired dataset was finally formed.

Unlike actual clinical data, the text content of the basic knowledge dataset is more standardized and consistent, and the question-answer pairs can cover the mainstream preoperative, intraoperative, and postoperative knowledge points of cataract surgery. This injects a solid medical foundation and professional expression ability into the model. The introduction of the multi-agent process ensures the relevance, consistency, and traceability of the generated questions and answers, and effectively avoids subjective bias or omissions in the manual sorting process.

The experimental results show that with the SFT pre-training of this part of basic knowledge, the model shows stronger medical concept understanding and medical terminology restoration capabilities in subsequent chain reasoning (CoT) and clinical recommendation tasks, and improves the overall generation quality indicators (such as BLEU, BERTScore, etc.). This shows that in the implementation of the large model medical field, the systematic and standardized injection of basic knowledge is a key link to improve the performance of downstream tasks.

\end{document}